\documentclass[a4paper,fleqn]{cas-dc}

\usepackage[numbers]{natbib}
\usepackage{bm}
\usepackage{float}

\newcommand{\nik}[1]{\textcolor{black}{#1}}

\def\tsc#1{\csdef{#1}{\textsc{\lowercase{#1}}\xspace}}
\tsc{WGM}
\tsc{QE}
\tsc{EP}
\tsc{PMS}
\tsc{BEC}
\tsc{DE}
\begin{document}
\let\WriteBookmarks\relax
\def\floatpagepagefraction{1}
\def\textpagefraction{.001}
\shorttitle{Dynein-driven self-organization of the microtubules}
\shortauthors{N. Frolov et~al.}

\title [mode = title]{Dynein-driven self-organization of microtubules: An entropy- and network-based analysis}

\author[1]{Nikita Frolov}[orcid=0000-0002-2788-1907]
\cormark[1]
\ead{nikita.frolov@kuleuven.be}
\credit{Conceptualization, Methodology, Software, Formal analysis, Visualization, Writing -- Original Draft Preparation, Writing -- Review & Editing}

\author[2]{Bram Bijnens}[orcid=0000-0002-2186-5455]
\credit{Software, Formal analysis}

\author[1]{Daniel Ruiz-Reyn{\'e}s}[orcid=0000-0003-0085-8421]
\credit{Conceptualization, Writing -- Original Draft Preparation}

\author[1]{Lendert Gelens}[orcid=0000-0001-7290-9561]
\cormark[2]
\ead{lendert.gelens@kuleuven.be}
\credit{Conceptualization, Methodology, Funding Acquisition, Supervision, Writing -- Review & Editing}

\address[1]{Laboratory of Dynamics in Biological Systems, Department of Cellular and Molecular Medicine, KU Leuven, 3000, Leuven, Belgium}
\address[2]{The Institute for Materials Research (imo-imomec), University of Hasselt, 3590, Diepenbeek, Belgium}

\cortext[cor1]{Corresponding authors}

\begin{abstract}
Microtubules self-organize to form part of the cellular cytoskeleton. They give cells their shape and play a crucial role in cell division and intracellular transport. Strikingly, microtubules driven by motor proteins reorganize into stable mitotic/meiotic spindles with high spatial and temporal precision during successive cell division cycles. Although the topic has been extensively studied, the question remains: What defines such microtubule networks' spatial order and robustness? Here, we aim to approach this problem by analyzing a simplified computational model of radial microtubule self-organization driven by a single type of motor protein -- dyneins. We establish that the spatial order of the steady-state pattern is likely associated with the dynein-driven microtubule motility. At the same time, the structure of the microtubule network is likely linked to its connectivity at the beginning of self-organization. Using the continuous variation of dynein concentration, we reveal hysteresis in microtubule self-organization, ensuring the stability of radial filament structures.
\end{abstract}

\begin{graphicalabstract}
\includegraphics[scale=0.3]{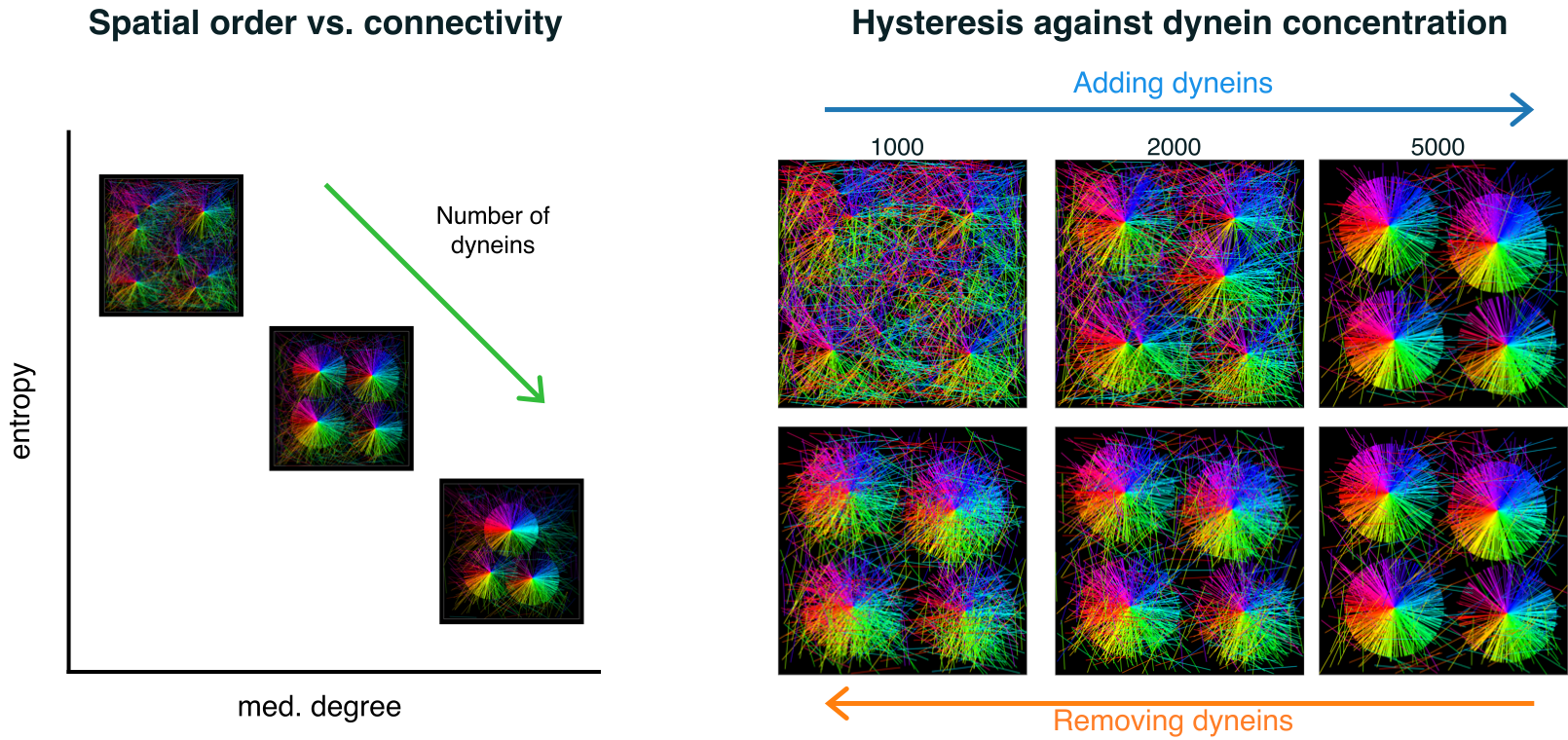}
\end{graphicalabstract}

\begin{highlights}
\item We study a simplified computational model for the self-organization of microtubules driven by dynein motor proteins.
\item  We demonstrate that the order of spatial organization increases progressively with dynein quantity, yet the spatial pattern changes abruptly.
%We show that the order of spatial organization grows continuously with the number of dyneins, but the spatial pattern change is discontinuous.
\item We deduce that while the former is likely associated with dynein-driven microtubule motility, the latter is probably explained by the changes in microtubule connectivity.
\item Using forward and backward continuations of the dyneins number, we show the hysteresis of the microtubules' radial arrangement.
\end{highlights}

\begin{keywords}
Microtubule network \sep Entropy \sep Pattern Formation \sep Agent-based modeling
\end{keywords}

\maketitle

\section{Introduction}

Nature exploits fascinating self-organization principles to provide the mechanisms of life on different scales. Just as the maintenance of an ecosystem depends on populations of species cooperating or competing for shared resources, life at the cellular level is determined by the arrangement and interaction between its constituents, such as proteins, polymers, and organelles~\cite{halatek2018self}. On a cellular level, disturbance of self-organization might lead to anomalies in morphology, spatial scaling, and temporal coordination, which can result in severe developmental disorders or cancer~\cite{zink2004nuclear,jevtic2019altering}.

Special biopolymers called \textit{microtubules} play a vital role in cellular organization. A microtubule is a fiber that dynamically alters its length, i.e., it grows or shrinks, through polymerization and depolymerization following the cell cycle progression~\cite{howard2007microtubule}. This process involves attaching or detaching the microtubule's building blocks -- tubulin dimers -- onto/from the microtubule ends. The one end that grows faster is known as a plus-end, and the opposite one is a minus-end. Microtubules are set in motion by bound motor proteins of two types: \textit{kinesins} walking towards the microtubule's plus-end and \textit{dyneins} moving in the opposite direction. This ensures the interaction of microtubules with other fibers, their collective motion, and organization into spatial structures~\cite{howard2009mechanical}. A recent theoretical study has shown that microtubule length regulation and motor-driven movement of the microtubules promote pattern formation in polymer-motor mixtures~\cite{striebel2022length}.

\begin{figure*}[!t]
	\centerline{\includegraphics[scale=0.2]{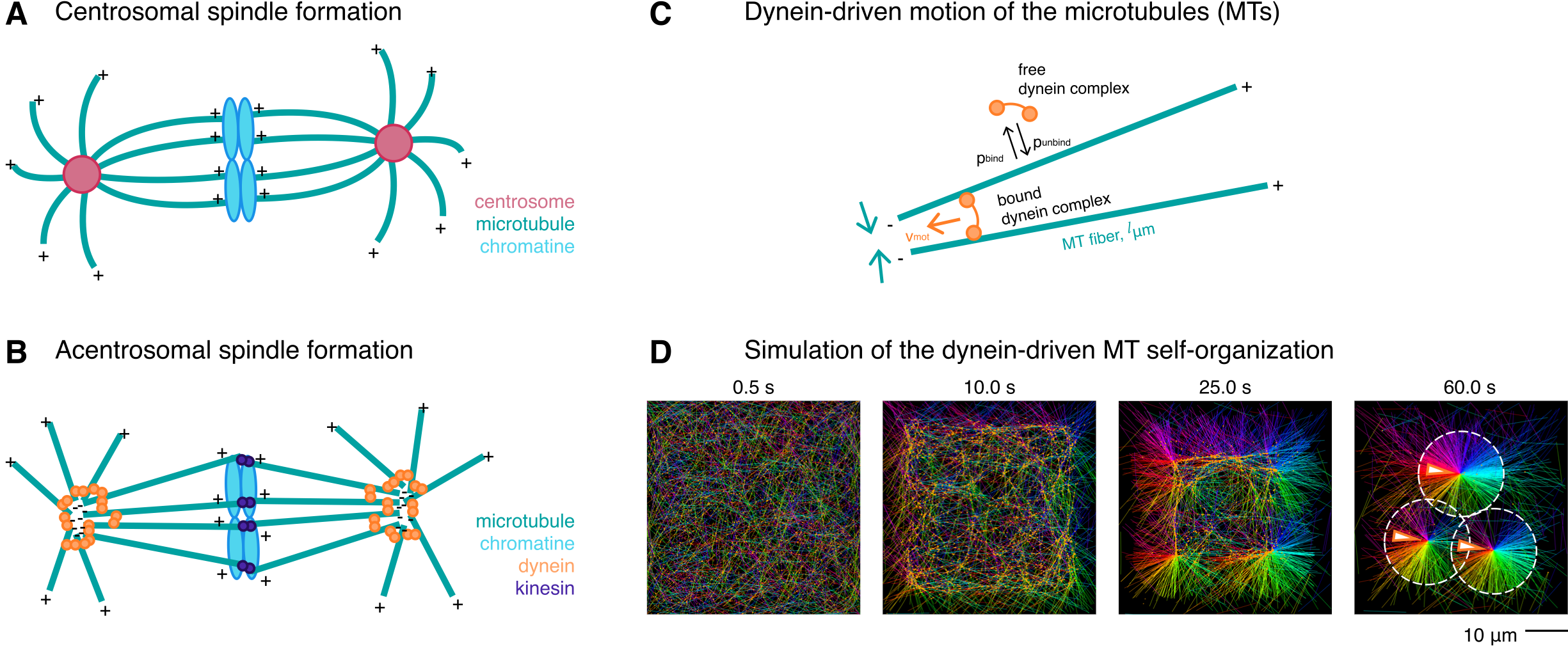}}
	\caption{\textbf{Self-organization of the microtubules into spindles and asters.} Microtubule asters are the critical structural components of the mitotic and meiotic spindles forming in centrosomal (\textbf{A}) and acentrosomal manner (\textbf{B}), respectively. \textbf{C.} Schematic of a minimal model of the microtubule-dynein interaction used in this study. \textbf{D.} Simulation of the dynein-driven self-organized microtubule network at $N_{\text{MT}=2000}$ and $N_{\text{mot}}=4000$. Snapshots are generated using the Cytosim; colors indicate microtubule orientation. Orange dots represent dynein complexes. Orange arrowheads indicate the position of dynein loci, and white dashed circles show the boundaries of separate microtubule asters.}\label{fig:1}
\end{figure*}

Due to the above-mentioned dynamical properties and ability to self-organize in space and time, microtubule networks are critical components of the cytoskeleton -- a scaffolding of a living cell ~\cite{fletcher2010cell,shamipour2021cytoplasm}. Strikingly, recent works reveal that the microtubules can organize into cell-like compartments even in a cell-free extract, i.e., a homogenized mixture of biochemical constituents of crushed living cells~\cite{cheng2019spontaneous,mitchison2021self,gires2023exploring}. Besides, microtubules play a significant role in cell division -- they accurately assemble into bipolar spindles that distribute replicated DNA among two daughter cells by pulling sister chromatids apart. In \textit{mitotic cells}, spindle poles are organized by star-like microtubule structures -- asters -- nucleated at the centrosomes and fixed by molecular motors (Fig.~\ref{fig:1} A). To date, a broad number of experimental studies explored biological mechanisms and properties of mitotic spindle formation \textit{in vivo} and \textit{in vitro}~\cite{sawin1992mitotic, heald1997spindle, gadde2004mechanisms, walczak2008mechanisms, good2013cytoplasmic, wilbur2013mitotic}. Notably, the centrosomes are absent in \textit{meiotic cells}~\cite{heald1996self, walczak1998model}, but opposed motors can still arrange bipolar spindles (Fig.~\ref{fig:1} B), revealing that the bipolarity is an essential property of self-organization in the motor-filament mixtures.

Deeper insight into the self-organization of microtubules and motors can also be gained from theoretical and computational models. Significant contributions in this direction were made by F. N{\'e}d{\'e}lec and colleagues, who developed the simulation software Cytosim~\cite{nedelec2007collective,gitlabCytosim} and used it to analyze hidden aspects of complex microtubule-motor interactions behind the filament organization~\cite{nedelec1997self,surrey2001physical,nedelec2001dynamic,karsenti2006modelling,loughlin2010computational,lacroix2018microtubule}. Remarkably, they have observed the emergence of particular microtubule structures such as asters and vortices, and they explained physical principles of their formation~\cite{surrey2001physical}. Later works have explored the stability of such patterns using nonlinear analysis and dynamical systems approach~\cite{lee2001macroscopic,sankararaman2004self,aranson2006theory}. Recently, Torisawa \textit{et al.} have demonstrated both experimentally and theoretically a diverse spatio-temporal patterning of microtubule fibers ranging from densely networked to isolated asters~\cite{torisawa2016spontaneous}.

Based on the importance of ordered microtubule arrangement for proper cellular structure and functioning and given a variety of spatio-temporal structures they can self-organize into, the question arises: which property of the microtubule-motor interaction defines the spatial order and stability of such a system? In this paper, we aim to address this issue in a simplified computational model of the microtubule-motor interaction, in which only one type of motor protein -- dynein -- drives the microtubule fibers. Firstly, we introduce the quantification of spatial order of microtubule arrangement using entropy -- an established measure of the order of the macroscopic states in statistical thermodynamics and complex systems science~\cite{mitchell2009complexity,thurner2018introduction}. We test how the order and structure of a spatial pattern change under the variation of the number of motor proteins. Secondly, we assess the association between the dynein-mediated motility and connectivity of the microtubule fibers. Consistent with previous studies, we show that an increased dynein concentration leads to a higher-order patterning in which microtubules arrange in star-like structures -- asters. Our results indicate that the properties of dynein-mediated microtubule connectivity likely determine the number of asters and their spatial arrangement. Finally, we address the stability of aster formation against perturbation of dynein concentration by analyzing the microtubule network self-organization in the forward and backward continuations. We reveal that the microtubule-motor mixture exhibits hysteresis under variation in the number of dyneins.

The paper is organized as follows. Section~\ref{sec:Model} describes the agent-based computational model of microtubule-motor interactions, its configuration, and its limitations. Section~\ref{sec:Spatial} explains the evaluation of the spatial order of the microtubule structures. In Section~\ref{sec:Results}, we present the study's main results and put them into the context of current literature. Section~\ref{sec:Concl} summarizes this work and highlights open questions.

\section{Modeling microtubule-motor mixtures}
\label{sec:Model}

Agent-based models provide an accurate mathematical description of complex interactions in multi-component systems. Individual agents follow straightforward rules (usually simpler than global behavior), making interpreting and controlling individual kinetic parameters easier. In this work, we perform agent-based modeling of the microtubule-motor interactions using Cytosim~\cite{nedelec2007collective,gitlabCytosim}, an open-source cytoskeleton simulation tool.

In Cytosim, microtubules and molecular motors are assumed to float around in liquid suspension, e.g., the cytoplasm in a cell. While motor proteins are considered small particles, larger objects like microtubule fibers are represented as sets of particles (points) connected by non-extensible rod segments of a fixed length $l_{\text{seg}}$. If dynamic instability of the microtubules is taken into account, points are added (removed) at the end of the fiber at rate $k_g$ ($k_s$) to model the growth (shrinkage) of the microtubule. Otherwise, fibers are considered to have a fixed length $l$.

Following Ref.~\cite{nedelec2007collective}, the motion of particles, both motor proteins and those of the fibers, obeys the Langevin equation in the form:
\begin{equation}
	\label{eq:Langevin}
	\frac{d \bm{r}}{dt} = \bm{B}(t) + \mu \bm{F}_{\text{ext}}(\bm{r},t),
\end{equation}
where $\bm{r}$ is a particle's position, $\bm{B}$ is a sum of the random collisions resulting in Brownian motion, $\mu$ is a particle's mobility, and $\bm{F}_{\text{ext}}$ accounts for the force acting on a particle at time $t$. For $N$ particles, Cytosim numerically solves the system of $N$ differential equations~(\ref{eq:Langevin}).

\begin{figure*}[!t]
	\centerline{\includegraphics[scale=0.20]{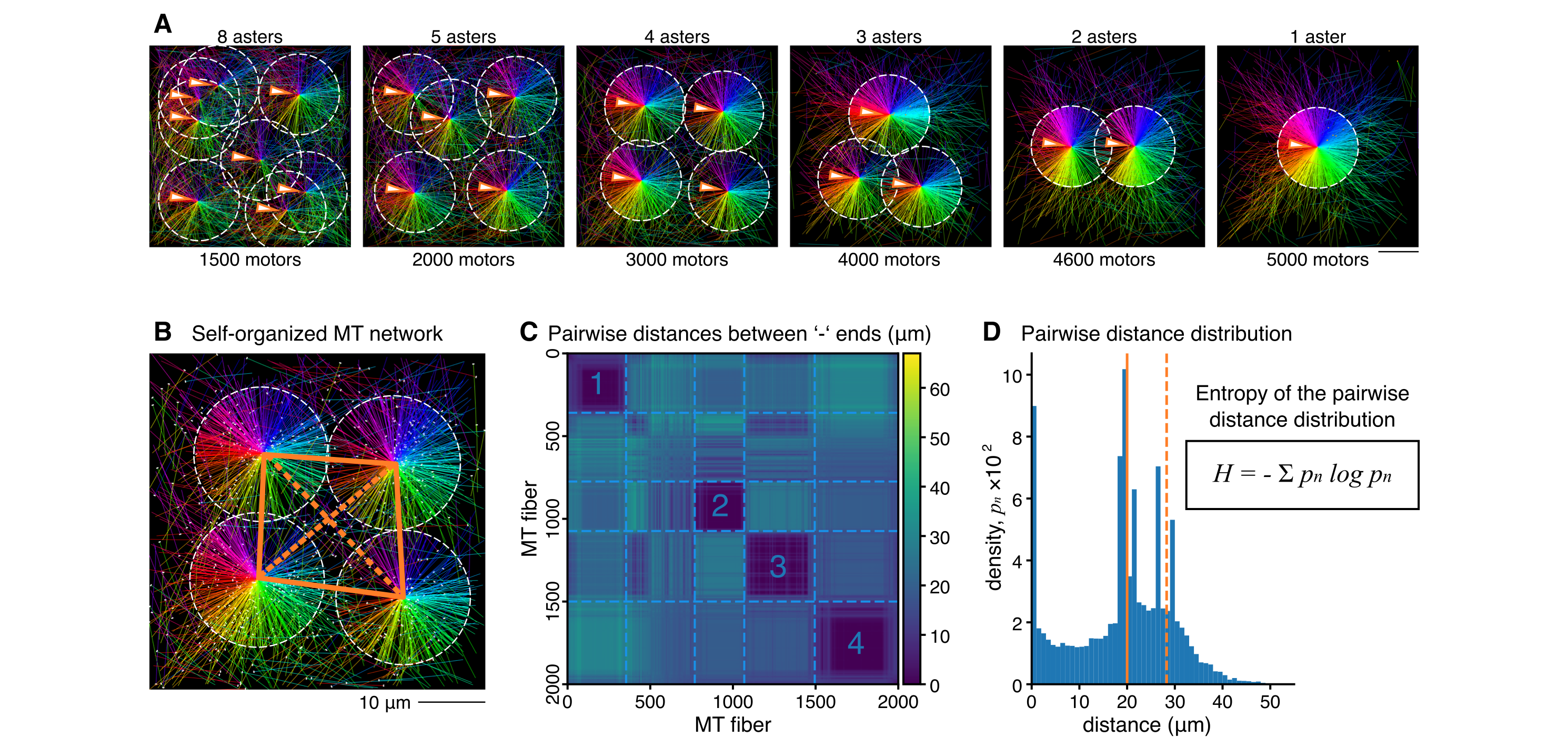}}
	\caption{\textbf{Microtubule self-organization under variation of the number of dyneins. A.} Evolution of the self-organized microtubule structure under an increase in the number of dynein complexes as captured by the end step of numerical simulation. Snapshots are generated using the Cytosim; colors indicate microtubule orientation. \textbf{B-D} Quantifying spatial order of the microtubule structures: (\textbf{B}) extracting locations of the microtubule fibers' minus-ends (white dots); (\textbf{C}) computing their pairwise distances; (\textbf{D}) constructing the pairwise distance distribution and evaluating its entropy. Solid and dashed orange lines in \textbf{B} and \textbf{D} indicate characteristic distances between neighboring and opposite asters, respectively.}\label{fig:2}
\end{figure*}

The motor proteins floating near a microtubule can bind to it with some probability. Attachment at rate $p_{\text{bind}}$ occurs on the closest site of the fiber segment when the motor is in the $\varepsilon$-vicinity of the fiber. While bound to the microtubule, the motor protein moves along the filament with effective velocity $v_{\text{mot}}=v_{0}(1-f/f_{\text{stall}})$, where $f$ is a load of the motor, $f_{\text{stall}}$ is a stall force, and $v_{0}$ is the motor's unloaded velocity. The sign of $v_{0}$ defines the direction of the motor's motion: positive $v_{0}$ determines the plus-end orientation, and negative $v_{0}$ -- minus-end one. Additionally, the motor can unbind from the microtubule at a load-dependent rate $p_{\text{unbind}}=p_0 \text{exp}(f/f_{\text{bind}})$, where $p_0$ is an unloaded unbinding rate and $f_{\text{bind}}$ is a binding force of the motor. A motor complex consisting of two motor proteins can bind to two fibers, thus creating a crosslink between them and generating forces that determine a fiber's motion and self-organization (Fig.~\ref{fig:1}C).

\textit{Assumptions}. In the current study, we make several assumptions to distill the effect of motor proteins on microtubule dynamics:
\begin{itemize}
\item we assume that only one type of molecular motors -- dyneins (negative $v_{0}$) -- are present in the system and set the microtubules in motion;
\item in large part of the simulations, we assume that the microtubules are of a fixed length $l$, neglecting their growth and shrinkage; some simulations account for the dynamic instability of the microtubules;
\item we consider a 2D model assuming that the microtubules and motors float in an infinitely thin square chamber of cytoplasm with a side length $l_{\text{ch}}>l$.
\end{itemize}

Simulation parameters are provided in Appendix~\ref{app:A}. In this study, we were focused on phenomenological modeling and sacrificed biological plausibility for faster simulations. The main parameters were selected according to Refs.~\cite{nedelec1997self,surrey2001physical}. The simulation starts with a random distribution of $N_{\text{mot}}$ dynein complexes over the chamber space at $t=0$. The positions and orientations of $N_{\text{MT}}$ microtubule filaments are also initiated randomly. We simulate 100s of the dynein-driven self-organization of the microtubules with the time step of $\Delta t=0.02$s (5000 iterations). Due to the stochastic nature of the simulations, we conducted $n=5$ realizations and averaged the results accordingly. An example of the simulation for $N_{\text{MT}}=2000$ and $N_{\text{mot}}=4000$ is presented in Fig.~\ref{fig:1}D.

\section{Spatial order of the microtubule structures}
\label{sec:Spatial}
Microtubules, being crosslinked and set into collective motion by the dynein complexes, self-organize to specific star-like structures -- asters -- with fibers' minus-ends pointing towards their centers (the process is also known as end-clustering). Varying the number of motor proteins, one can switch from a weakly organized network of poorly concentrated asters to a giant aster accumulating all available microtubule filaments (Fig.~\ref{fig:2}A). We quantify the macroscopic state through the entropy of minus-ends positions to gain insight into the transition from a disordered to an ordered steady state of the microtubule-motor network.

\textit{Minus-ends positions.} As mentioned before, with an increasing number of dyneins $N_{\text{mot}}$, the spatial pattern changes from lower-order states with randomly distributed and randomly oriented microtubule filaments to the formation of asters with the fibers' minus-ends densely concentrated near their centers -- states of a higher order. In this context, it is logical to describe the order of the spatial patterns in terms of the minus-ends distribution. Plots in Fig.~\ref{fig:2}B-D represent a step-by-step entropy-based quantification of the macroscopic state of the microtubule network. Fig.~\ref{fig:2}B shows an exemplary distribution of the filaments (green lines) with a few asters formed in a steady state at $N_{\text{mot}}=3000$. Most filaments are arranged in the asters, and some freely float in the cytoplasm. The blue dots indicate the positions of the minus-ends, most of which, as expected, are concentrated around the centers of respective asters, resulting in almost 0 $\mu$m distance between the neighboring ends. Other spatial scales emerge from the distance between the formed asters as indicated by solid and dashed orange lines in Fig.~\ref{fig:2}B.

\textit{Pairwise distances.} Based on the discussion above, it is reasonable to assume that the pairwise distances between the fibers' minus-ends contain information on how ordered the microtubule network is. We compute the pairwise distance matrix $D$ as:
\begin{equation}
	D_{ij}	= ||\bm{r}^-_{i} - \bm{r}^-_{j}||,
\end{equation}
where $\bm{r}^-_i$ is a position of the minus-end of the $i^{th}$ microtubule fiber, and $||\bullet||$ is a Euclidean norm. Fig.~\ref{fig:2}C displays the matrix $D$ corresponding to the spatial pattern in the left panel. The matrix $D$ reflects the spatial structure of the microtubule pattern, i.e., four clusters of closely located minus-ends (the deep blue squares along the main diagonal enumerated 1, 2, 3, and 4) associated with respective asters. Alongside these clusters, the matrix isolates the group of freely distributed filaments (the region between squares 1 and 2).

\textit{Pairwise distance distribution.} Here, we describe the pairwise distance distribution and use it to quantify how ordered the microtubule structure is. Fig.~\ref{fig:2}D shows a respective probability density function (PDF) of the pairwise distances $D_{ij}$ (bin width of 1 $\mu$m). As expected, the distribution is multimodal with the local peaks around characteristic spatial scales: (i) 0-scale, i.e., the minus-ends concentrated near the asters' centers, and (ii) the scale of the distance between asters. Notably, there are few local peaks around the second mode of the distribution. One corresponds to the minimal distance between neighboring non-overlapping asters, roughly $2l=20$ $\mu$m (solid vertical line) -- the closest distance for two asters to stay in equilibrium and not collapse in a single one. The other reflects the distance between the opposite asters (dashed vertical line $\approx2\sqrt{2}l=28.28$ $\mu$m, dashed diagonal lines in the left panel). Finally, we evaluate the macroscopic steady state using Shannon's definition of entropy~\cite{shannon1948mathematical}:
\begin{equation}
	H = -\sum{p_n \log{p_n}},
\end{equation}
where $p_n$ is a probability density of the $n^{th}$ bin. Indeed, with such a definition, a completely disordered state having a normal distribution of pairwise distances and large variance is characterized by the highest entropy. On the contrary, dynein-driven self-organization facilitates aster nucleation and diminishes the pairwise distance variance between microtubules' minus-ends. It results in sharp peaks in the pairwise distance distribution, i.e., an entropy reduction.

\section{Results and Discussion}
\label{sec:Results}

\subsection*{Dynein concentration affects the spatial order of the microtubule arrangement}

\begin{figure*}[!t]
	\centerline{\includegraphics[scale=0.20]{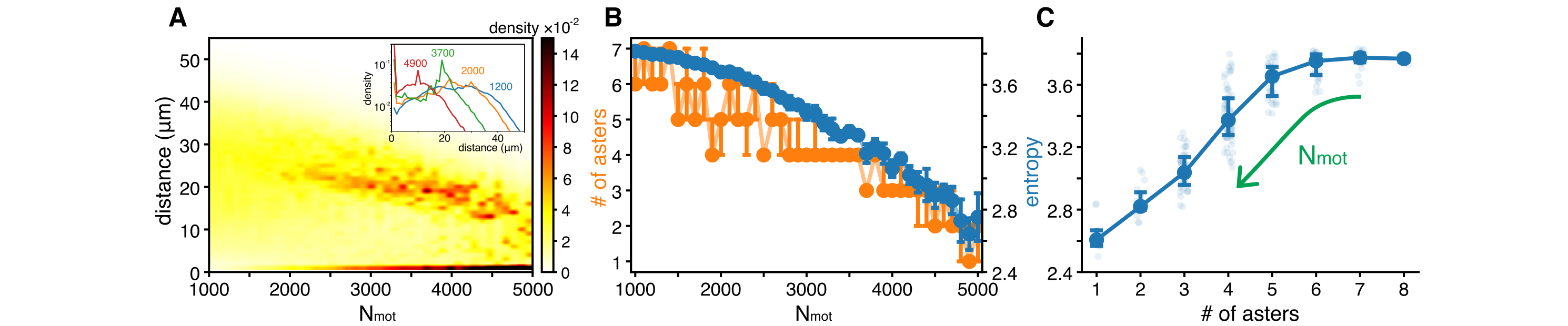}}
	\caption{\textbf{Characterizing spatial order of the steady state microtubule pattern under variation of the number of dyneins. A.} Evolution of the pairwise distance distribution with increasing total dynein concentration. The 2D plot is averaged across $n=5$ realizations. \textbf{B.} The median number of asters (orange circles) and mean entropy of the pattern (blue circles) versus the total dynein concentration. Error bars indicate interquartile range (IQR) and standard deviation (SD). \textbf{C.} The relationship between entropy and the number of asters. Non-transparent circles show the median entropy estimate for a given number of asters. The green arrow indicates the direction of $N_{\text{mot}}$ growth.}\label{fig:3}
\end{figure*}

We examine how the introduced quantification of spatial order applies to the microtubule patterns formed at different numbers of the dynein complexes $N_{\text{mot}}$. Fig.~\ref{fig:3}A shows the evolution of the pairwise distance distribution for an increasing number of motor proteins. One can see that at $N_{\text{mot}}\lessapprox2000$, the distribution of pairwise distances is the flattest, reflecting a disordered steady state with poorly nucleated microtubule asters. Increasing the number of molecular motors up to $\approx3500$ leads to the emergence and rise of the peaks in the distribution around 0, 20, and 30 $\mu$m, indicative of an arrangement of densely concentrated asters into well-organized square-shaped structures (see the inset in Fig.~\ref{fig:3}A and Fig.~\ref{fig:2}B, D). In the range $N_{\text{mot}} \approx 3500 - 4600$, only two peaks remain in a distribution, around 0 and $2l$ $\mu$m, and their magnitude increases considerably compared to lower dynein concentrations. It indicates breaking a square state symmetry, transforming the spatial pattern into a triangular shape, and then further into two isolated asters. Finally, at $N_{\text{mot}}>4600$, microtubule fibers are accumulated in a single highly nucleated aster as displayed by a prominent sharp peak around 0 $\mu$m in the pairwise distance distribution. 

Fig.~\ref{fig:3}B displays the entropy of the pairwise distance distribution (blue circles) alongside the number of asters versus the dynein concentration (orange circles). The variation of the number of asters with an increasing $N_{\text{mot}}$ is stepwise and discontinuous, having multiple plateaus. For example, the region of $N_{\text{mot}}\leq 2700$ is characterized by the highest number of asters ($\geq$5) as well as its high variance, however at $N_{\text{mot}}\in(2700,3700)$, the microtubule-motor mixture robustly converges to a four aster pattern. Such an alternation of the regions of stability and variability of the steady-state pattern suggests the multistability of the dynein-driven self-organization of the microtubules (discussed further). In turn, entropy continuously decreases with the dynein concentration growth. It reflects a gradual increase of the spatial order while adding more molecular motors.
	
Establishing an explicit association between entropy and the number of asters (Fig.~\ref{fig:3}C), one can see that the patterns containing six or more asters are almost equally disordered as provided by saturation of entropy around the value of 3.8. Conversely, the least complex states with one highly concentrated aster are characterized by an entropy of about 2.6. Gradual increase of entropy between these values, i.e., for 2, 3, 4, and 5 asters, suggests the complication of an arrangement of the asters into a global spatial pattern. While for 3 and 4 asters, a nonequilibrium steady-state reaches an ordered spatially-homogeneous structure (triangle and square, respectively), a 5-aster-pattern is of higher spatial heterogeneity due to partial overlapping of a few asters (see examples in Fig.~\ref{fig:2}A).

\begin{figure*}[!t]
	\centerline{\includegraphics[scale=0.20]{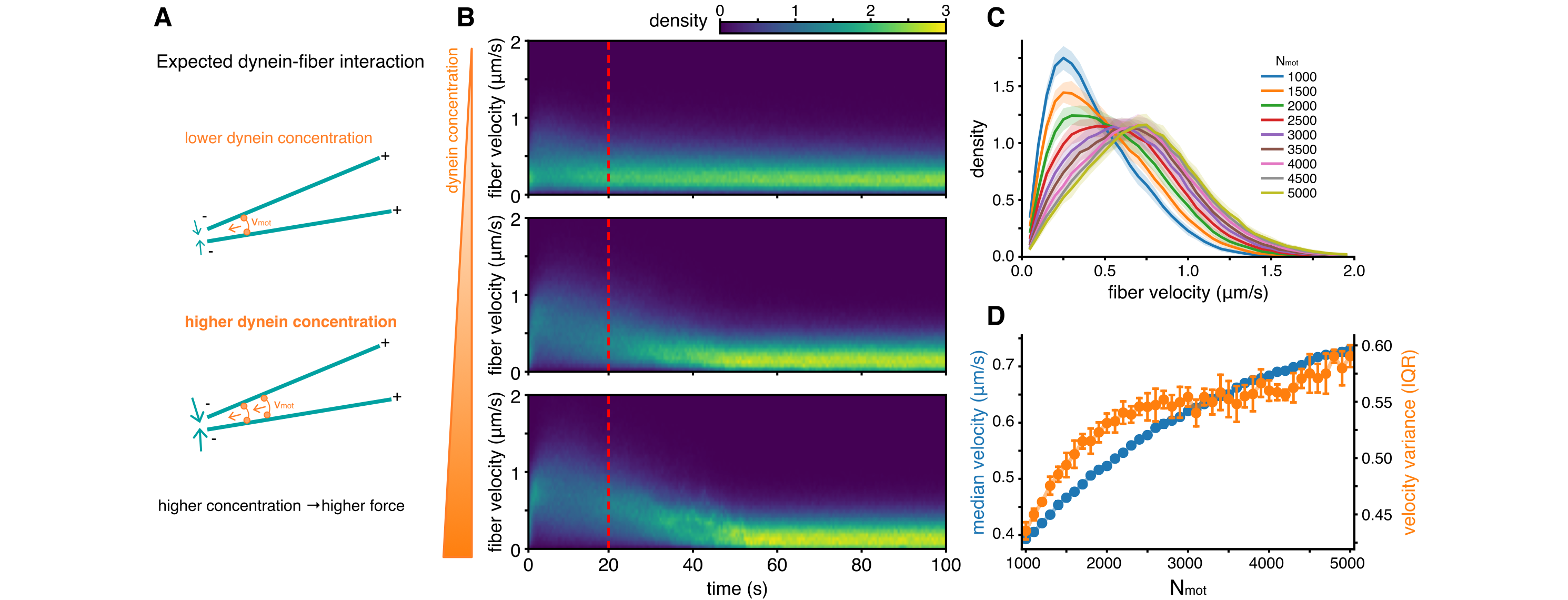}}
	\caption{\textbf{A Higher number of dynein complexes increases the motility of the microtubule fibers at the beginning of self-organization. A.} Schematic of the expected effect of the dynein concentration on microtubule dynamics. The size of the green arrows reflects the magnitude of the force exerted by the dyneins on the microtubules. \textbf{B.} Evolution of the microtubule fiber velocity distribution in the course of self-organization under an increase of the number of dyneins $N_{\text{mot}}$ (top-down). The 2D plots are averaged across $n=5$ realizations.\textbf{C.} Averaged fiber velocity distribution at the beginning of self-organization (indicated by the vertical dashed line in \textbf{B}) for different numbers of dynein complexes $N_{\text{mot}}$ (mean$\pm$SD across $n=5$ realizations). \textbf{D.} Median value (blue) and inter-quartile range (IQR, orange) of the fiber velocity distribution from the subplot \textbf{C} versus the number of dynein complexes $N_{\text{mot}}$ (mean$\pm$SD across $n=5$ realizations).}\label{fig:4}
\end{figure*}

As the resulting spatial organization of the microtubule-motor mixture is constrained by the dimensions of the chamber, we have conducted a similar analysis in a chamber of a larger size, 75$\times$75 $\mu$m$^2$ (Appendix~\ref{app:B}, Fig.~\ref{fig:B1}). For the comparison purpose, we kept the same concentration of the microtubules and molecular motors (See Table~\ref{tab: parameters_large} for the simulation parameters). Our results show that in a larger chamber, the microtubules self-organize into more spatially heterogeneous structures, which is allowed by larger spatial scales of the system (Fig.~\ref{fig:B1} A), as indicated by an elevated level of entropy and a larger number of asters (Fig.~\ref{fig:B1} B, C). However, the dependence of both of these quantifiers on motor concentration does not fundamentally change compared to a smaller chamber considered before. In larger and smaller chambers, higher entropy corresponds to a greater number of asters and vice versa (Fig.~\ref{fig:B1} D). \nik{To compensate for the difference in chamber size, we additionally considered normalized entropy $H/\log(N_{MT})$ (Fig.~\ref{fig:B2}). In a strongly disorganized state, i.e., at the lowest considered motor density, the normalized entropy is around 0.5 for both chambers. In a smaller chamber, normalized entropy reduces much faster under an increase in motor density than in the larger chamber. Interestingly, the level of normalized entropy for 1 isolated aster in the smaller chamber roughly corresponds to one for 2-3 asters in a larger chamber. One may conclude that an increase in a chamber size does not simply scale the patterning of the microtubule-motor mixture. On the contrary, it introduces additional complexity and heterogeneity to the self-organization by less constraining the interaction space.}

Our computational results emphasize that the dyneins help maintain microtubule networks' stability and spatial order. Similarly, previous works have shown the transition from globally connected (disordered) microtubules to sparse concentrated metastructures by changing dynein concentration~\cite{nedelec1997self,surrey2001physical}. A more recent study by Torisawa \textit{et al.}~\cite{torisawa2016spontaneous} has identified a very similar transition from static to globally and locally contractile microtubule networks under the variation of motor/filament ratios. In Ref.~\cite{lemma2022active}, Lemma et al. construct a more comprehensive map of nonequilibrium states in the microtubule-kinesin mixture ranging from asters and contracting gel to active foam and extensile fluid depending on the concentration of motors, tubulin, and particles promoting attractive forces between filaments. Emphasizing the role of asters as critical structural elements of mitotic spindles, Hueschen et al. have reported that dynein-deficient mitotic cells exhibit disordered and temporally unstable spindles~\cite{hueschen2019microtubule}.

\subsection*{Dynein concentration affects the motility of microtubules at the beginning of self-organization}

To better understand how the dynein concentration affects the spatial order in this system, we analyzed the motility of the microtubules under variation of the number of dyneins $N_{\text{mot}}$. At higher dynein concentrations, more molecular motors can attach to the pairs of microtubules and pull them towards each other with greater force (Fig.~\ref{fig:4}A). One can expect that higher dynein concentration would increase the motility of the microtubules. 

To assess this effect, we analyzed the distribution of the fibers' velocity in the course of self-organization at different $N_{\text{mot}}$ (Fig.~\ref{fig:4}B). The fiber's velocity was evaluated as a rate of its minus-end displacement. The top panel in Fig.~\ref{fig:4}B shows that at a small dynein concentration, molecular motors do not affect the microtubules' motility -- the velocity distribution does not change in time, and the system immediately finds itself in a disordered steady state. At higher dynein concentrations (middle and bottom panels in Fig.~\ref{fig:4}B), the motility of the microtubules rapidly increases at the beginning of self-organization. It is reflected in the elevated median and the variation of the fibers' velocity shortly after the start of the simulation compared to a steady-state distribution. Afterward, it gradually relaxes until it reaches a steady state.

Considering the velocity distribution averaged across the beginning of self-organization (first 20s, limited by the dashed lines in Fig.~\ref{fig:4}B), one can see its systematic shift towards higher average velocities as $N_{\text{mot}}$ increases (Fig.~\ref{fig:4}C). Moreover, it becomes more positively skewed since the variance grows faster than the median value (Fig.~\ref{fig:4}D). The progressive distribution asymmetry can be interpreted by the growth of dynein-driven motility that considerably speeds up fibers at the periphery due to higher global contractility (discussed further), while those in the center immediately group in the asters. It results in centered and much densely nucleated asters (Fig.~\ref{fig:2}A).

Smooth growth of the median velocity of the microtubules with the dynein concentration (Fig.~\ref{fig:4}D) can potentially explain a continuous reduction of spatial entropy, i.e., higher motility of the microtubules promotes nucleation of more concentrated asters. However, this quantity is insufficient to describe a discontinuous transition in the number of asters and their spatial arrangement. 

\begin{figure*}[!t]
	\centerline{\includegraphics[scale=0.2]{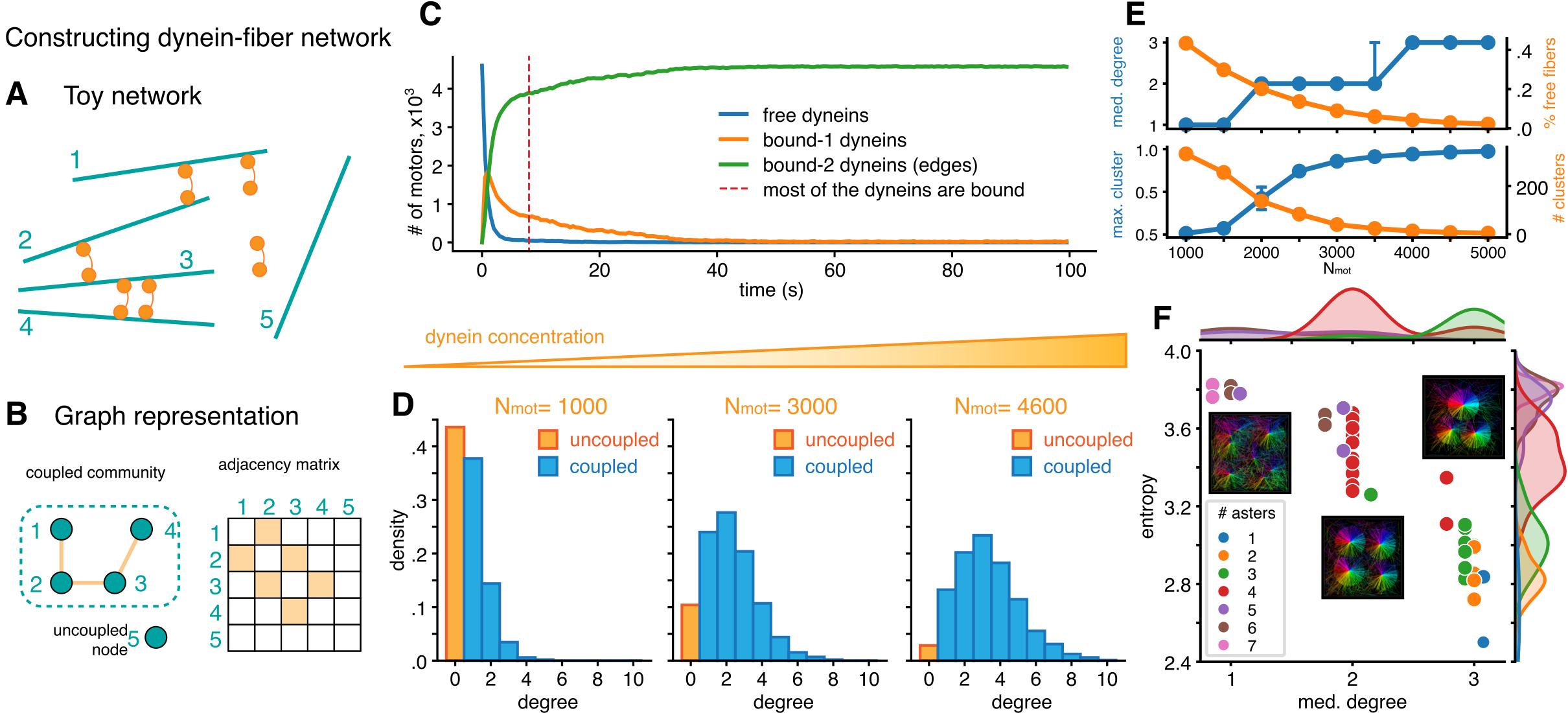}}
	\caption{\textbf{Higher dynein-mediated connectivity of microtubule fibers results in more ordered microtubule structures. A.} Constructing dynein-fiber network. Exemplary toy network consisting of fibers (green) coupled mechanically through the bound dyneins (orange). \textbf{B.} Representing dynein-fiber network as a graph. Here, fibers enumerated 1,2,...,5 are considered vertices, and the dynein complexes bound to two fibers (bound-2 dyneins) are considered mechanical edges. The resulting graph is associated with an adjacency matrix from which useful network metrics are assessed. \textbf{C.} Exemplary time traces of the number free (blue), bound-1 (orange), and bound-2 (green) dyneins in the course of self-organization at $N_{mot}=4000$. Vertical dashed lines indicate the moment most dyneins are bound to fibers. \textbf{D.} Degree distribution in the initial dynein-fiber network under increasing dynein concentration. \textbf{E.} Top: median degree (blue; median and IQR) and the number of uncoupled fibers (orange, mean$\pm$SD) versus the dynein concentration. Bottom: maximal cluster size (blue; mean$\pm$SD) and the number of clusters (orange, mean$\pm$SD) versus the dynein concentration. \textbf{F.} Association between the structural entropy of self-organized microtubule network and the median degree of initial dynein-fiber network. Data points are color-coded with the number of asters.}\label{fig:5}
\end{figure*}

\subsection*{Dynein-mediated connectivity of microtubules affects their spatial order}

Apart from the fibers' motility, dyneins mediate the connectivity of the microtubule network, another crucial factor affecting self-organization in motor-filament mixtures. Intuitively, a higher number of dynein promotes stronger connectivity of the microtubules. However, it is unclear if the dynein-regulated connectivity changes can better explain discontinuous transitions in the number of asters into which the microtubules arrange. If so, how should one measure connectivity?

Here, we employ graph theory to quantitatively describe the dynein-mediated connectivity of the microtubules from the simulated data~\cite{belmonte2017theory, eliaz2020insights}. To illustrate our approach, we use a toy dynein-fiber network displayed in Fig.~\ref{fig:5}A. We consider microtubules as vertices with dyneins mechanically connecting them. Dyneins, therefore, can be found in one of the three states: (\textit{i}) free dyneins, not bound to any microtubule; (\textit{ii}) bound-1 dyneins, attached to only one microtubule; (\textit{iii}) bound-2 dyneins, attached to two microtubules. Only the latter ones establish pairwise coupling of the filaments and, thus, are considered as edges. By defining vertices and edges in this way, we represent the toy dynein-fiber network as a graph with fibers 1--4 constituting a cluster and fiber 5 being an uncoupled or isolated vertex (Fig.~\ref{fig:5}B).

Using this graph representation of the dynein-fiber network, we seek to determine if individual microtubule connectivity, i.e., microscopic properties, is associated with the global network's emergent mesoscale structures (microtubule clusters and asters). Considering the adjacency matrix of the graph $\mathcal{A}_{ij}$ binary, we use a measure of vertex degree (the number of adjacent edges) defined as $k_i = \sum_{j=1}^{N_{\text{MT}}}\mathcal{A}_{ij}$, $i\in[1, N_{\text{MT}}]$ as the microscopic property of the microtubule-motor network. In this notation, degree $k_i$ indicates the number of microtubules connected to the $i^{th}$ microtubules through the dyneins, regardless of the number of dyneins binding them. We describe mesoscopic structures by finding all the dynein-fiber network clusters and computing the maximal cluster size and the number of clusters. Noteworthy, the measure of maximal cluster size was used to distinguish between the polar and nematic organization of the microtubule networks in Ref.~\cite{roostalu2018determinants}.

\begin{figure*}[!t]
	\centerline{\includegraphics[scale=0.18]{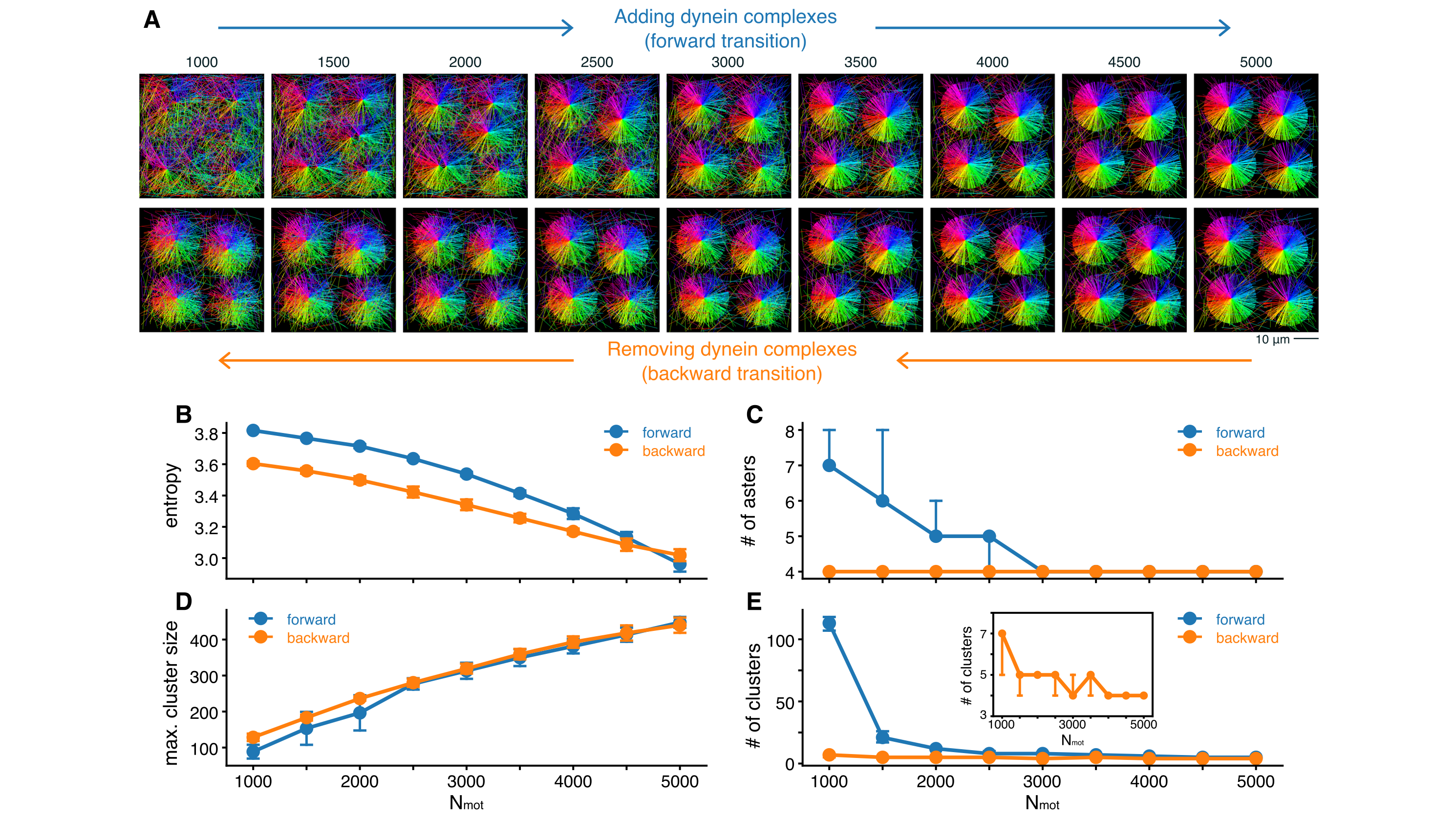}}
	\caption{\textbf{Self-organization of the microtubules exhibits hysteresis under a continuous variation of dynein concentration. A.} Evolution of a steady state in a microtubule-motor mixture. Top row: forward transition, i.e., continuously adding new dynein complexes into the chamber; bottom row: backward transition, i.e., continuously removing existing dynein complexes from the chamber. \textbf{B-E.} Dependence of the steady state quantifiers on the number of dynein complexes during forward (blue) and backward (orange) transitions: (\textbf{B}) entropy (mean$\pm$SD across $n=5$ realizations); (\textbf{C}) number of asters (median and IQR); (\textbf{D}) maximal cluster size (mean$\pm$SD); (\textbf{E}) number of clusters (mean$\pm$SD across $n=5$ realizations).}\label{fig:6}
\end{figure*}

Since the dyneins are randomly distributed in space and detached from the microtubules at the start of the simulation, it takes some time for the molecular motors to bind fibers and establish a connected dynein-fiber network (Fig.~\ref{fig:5} C for $N_{\text{mot}}=4000$). Thus, we analyze the network connectivity properties at the time indicated by the dashed vertical line in Fig.~\ref{fig:5}C when almost all dyneins are bound to the microtubules (the number of free dyneins is $<1$\%). Notably, it is associated with the highest motility of the microtubule fibers linking connectivity and contractility of the network (compare with Fig.~\ref{fig:4}B). Using the graph representation of the spatial dynein-fiber network at the beginning of self-organization, we evaluate distributions of the microtubules under the variation of $N_{\text{mot}}$ (Fig.~\ref{fig:5} D). Firstly, one can see that the amount of uncoupled fibers with $k_i=0$, orange bin in the distributions, makes up almost half of the fibers population at small $N_{\text{mot}}$ and continuously fades as the dynein concentration increases (Fig.~\ref{fig:5} D, E). Secondly, as expected, an increase in the number of dyneins shifts the degree distribution to higher values, i.e., microtubule fibers connect with a larger number of neighbors. Unlike the proportion of uncoupled fibers, the median degree exhibits a stepwise variation against $N_{\text{mot}}$ (Fig.~\ref{fig:5} E, top panel). On the mesoscopic level, the number of detected clusters gradually decreases and converges to the presence of 1 giant cluster with increasing dynein concentration. In contrast, the maximal size of the cluster starts growing at $N_{\text{mot}}=2000$ and saturates at the level of 1 corresponding to the total amount of microtubule filaments (Fig.~\ref{fig:5}E, bottom panel). \nik{In a larger chamber, this saturation is sufficiently delayed and takes place at a roughly 4-fold higher motor density compared to a considered chamber (Fig.~\ref{fig:B1}C).}

This mesoscopic description of the microtubule-motor network resembles the percolation transition~\cite{stauffer2018introduction}. The fibers tend to concentrate in a giant aster under the growing number of dyneins, which enhance network connectivity by crosslinking fibers and exerting contractile forces. Strikingly, the onset of the percolation transition corresponds to the increase of the median microtubule degree, establishing the link between microscopic fiber properties and the arrangement of the global network on the mesoscopic level. In addition, the variation of median degree agrees with the structural reorganization of the microtubule-motor network expressed in terms of the number of asters (Fig.~\ref{fig:3}B). Plotting the spatial entropy versus the median degree of the dynein-fiber network at respective $N_{\text{mot}}$, we achieve a good clustering of steady-state microtubule patterns in this parameter plane (Fig.~\ref{fig:5}F). Highly disordered states with many poorly nucleated asters are characterized by the median degree of 1, meaning that sparsely connected pairs of microtubules dominate the dynein-fiber network. Thus, such a fragmented network cannot be organized coherently. A median degree of 2, indicative of the arrangement of microtubules in larger communities through pairwise interactions, characterizes spatial structures with multiple prominent spatial scales (5 and 4 asters). Finally, single-scale structures, such as 3-, 2-, and 1-aster states, are characterized by a median degree of 3, implying the dominance of higher-order coupling and tighter connectivity and contractility in the dynein-fiber network. Thus, we linked microtubule connectivity with a steady-state structure by translating a dynein-fiber network into a graph and quantifying it through the degree distribution.

\begin{figure*}[!t]
	\centerline{\includegraphics[scale=0.20]{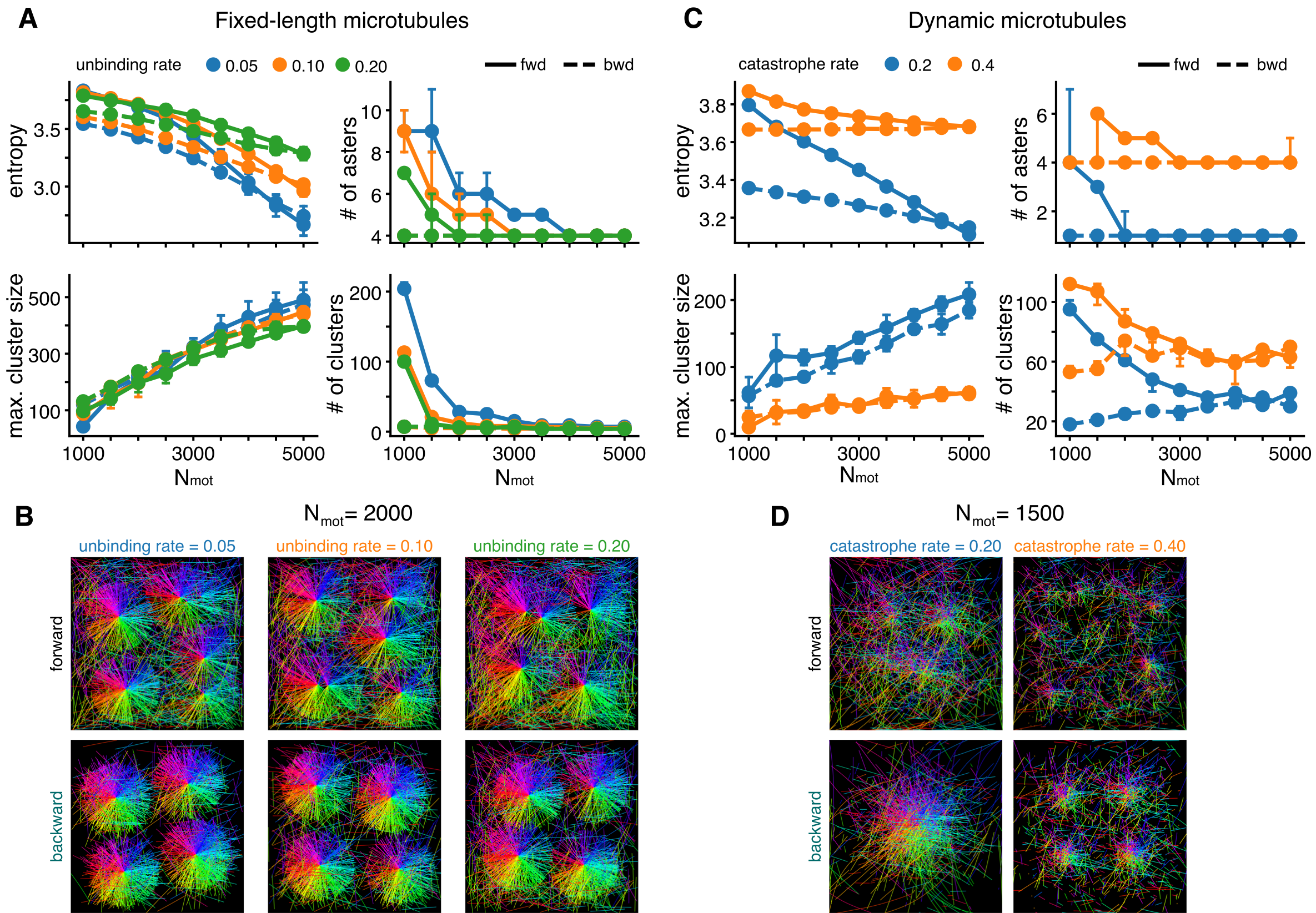}}
	\caption{\textbf{Hysteresis is preserved at different dynein unbinding rates and in dynamically unstable microtubules. A.} Dependence of the steady state quantifiers on the number of dynein complexes during forward (solid lines) and backward (dashed lines) transitions color-coded with the motor unbinding rate. \textbf{B.} Snapshots of the steady state in filament-motor mixture with static microtubules at the fixed number of motors $N_{\text{mot}}=2000$ but for different motor unbinding rates during forward (top row) and backward (bottom row) transitions. \textbf{C.} Dependence of the steady state quantifiers on the number of dynein complexes during forward (solid lines) and backward (dashed lines) transitions color-coded with the catastrophe rate. \textbf{D.} Snapshots of the steady state in filament-motor mixture with dynamic microtubules at the fixed number of motors $N_{\text{mot}}=1500$ but for different catastrophe rates during forward (top row) and backward (bottom row) transitions. \nik{In panels \textbf{A} and \textbf{C}, entropy and max cluster size are presented as mean$\pm$SD across $n=5$ realizations. The number of asters and the number of clusters are presented as median and IQR across $n=5$ realizations.}}\label{fig:7}
\end{figure*}

\subsection*{Self-organization of the microtubules exhibits hysteresis under a continuous variation of dynein concentration}

As mentioned before, variability of the number of asters in the microtubule network steady-state (See Fig.~\ref{fig:3}B) may be indicative of the multistability of self-organization, meaning that multiple steady-states are allowed at the same set of system parameters. Multistability is an essential feature of living systems that ensures the robustness of their steady states against internal or external perturbations~\cite{pisarchik2014control}. The multistable nature of the dynein-driven microtubule self-organization may explain the stability and reproducibility of meiotic and mitotic spindle formation during successive cell division cycles.

To address this issue in our model, we conducted an additional numerical simulation to compare dynein-driven microtubule self-organization in \textit{forward} and \textit{backward} continuations. To achieve forward transition, we first progressively increased the number of the dyneins in the chamber from 1000 to 5000, with the step of 500 reaching a steady-state pattern at each step. Each new portion of the dyneins was randomly distributed within the chamber. The backward transition was obtained by repeating the same iterative procedure but in the opposite direction, i.e., decreasing the number of motors from 5000 to 1000 by randomly removing 500 motors at each subsequent step. In this scenario, the multistable microtubule-motor mixture is expected to exhibit \textit{hysteresis}.

Fig.~\ref{fig:6} displays the evolution of the dynein-fiber system in the forward and backward directions. The steady-state snapshots in Fig.~\ref{fig:6}A indicate the presence of hysteresis. During forward continuation, the microtubule-motor mixture undergoes a percolation transition starting from a poorly organized and almost non-contractile network at $N_{\text{mot}}=1000$ to four highly nucleated non-overlapping asters due to enhanced connectivity and contractility produced by a growing number of dyneins. Continuous dynein removal during backward transition decreases network connectivity, thus releasing free fibers from the asters into the cytoplasm. At the same time, the 4-asters steady-state formed at higher dynein numbers stays largely non-disrupted at lower motor concentrations.

The steady state of the microtubule system during forward and backward continuation is evaluated using the entropy- and network-based quantifiers (Fig.~\ref{fig:6}B-E). Fig.~\ref{fig:6}B shows that the backward transition steady states exhibit a higher degree of spatial order (lower entropy) than ones of the forward transition. While the entropy variation is smooth, network-based features display the transition from the poorly contractile network to highly nucleated asters around 2500-3000 dyneins during the forward continuation. However, the backward continuation diagrams do not repeat the same transition, thus quantitatively proving the presence of hysteresis.

As described above, hysteresis reveals a remarkable property of the dynein-driven aster formation: once organized, the aster is not sensitive to small perturbations of the molecular motor activity manifesting stability of the radially arranged microtubules. This property is critical in the robustness and reproducibility of meiotic and mitotic spindle reorganization in repetitive cell division cycles. Strikingly, in their recent work\footnote{The paper by Scrofani et al. was published after the submission of the current manuscript}, Scrofani et al. observed \textit{in vitro} that the dynein inhibition does not collapse the pre-formed RanGTP aster~\cite{scrofani2024branched}. This result provides indirect but remarkable experimental support for the currently observed hysteresis in the dynein-driven microtubule aster organization.

\subsection*{Hysteresis is preserved at different dynein unbinding rates and in dynamic microtubules}

Lastly, we assess the influence of the motor and microtubule parameters on the hysteresis of dynein-driven aster organization. With this aim, we vary the dynein unbinding rate $p_0$ in a static microtubule scenario and separately analyze the hysteresis in the self-organization of dynamically unstable microtubules and the effect of catastrophe rate $k_{\text{cat}}$. Additionally, we test if the above-described microtubules' self-organization remains the same at a more realistic (slower) motor speed $v_0$.

Figs.~\ref{fig:7}A, B display the forward and backward continuations of static microtubules self-organization at different unbinding rates of motor proteins $p_0$. We compared the pattern formation process at $p_0=0.10$ (the reference value), $p_0=0.05$ (two times lower than the reference value), and $p_0=0.20$ (two times higher than the reference value). Fig.~\ref{fig:7}A shows that hysteresis is present at all considered values of $p_0$. Remarkably, a lower dynein unbinding rate $p_0=0.05$ (blue) leads to the steady states of higher spatial order (lower entropy) with stronger nucleated asters (higher max. cluster size). However, the threshold of percolation transition is shifted toward the higher values of dynein numbers. Alternatively, at a higher unbinding rate $p_0=0.20$ (green), the microtubule-motor network possesses a lower spatial order (higher entropy) with less nucleated asters (lower max. cluster size), and the percolation transition threshold is shifted to the lower number of motors. These observations imply that at a lower unbinding rate, the microtubule-motor network rewiring through the detaching dyneins is less frequent, which helps maintain the established radial clusters of microtubules during the backward transition but does not help accumulate new fibers into asters, thus delaying the forward transition onset.

Account of the microtubules' dynamical instability also resulted in hysteresis at the catastrophe rates of 0.2 and 0.4 (Fig.~\ref{fig:7}C, D). We adjusted the microtubules' growth parameters (Table~\ref{tab: parameters_dynamic}) to reach the averaged filament length of 2.5 $\mu$m at $k_{\text{cat}}=0.4$ and 5.0 $\mu$m at $k_{\text{cat}}=0.2$. At a lower catastrophe rate $k_{\text{cat}}=0.2$, the microtubule network undergoes a quick percolation transition to a giant aster. In comparison, short microtubules formed at a higher catastrophe rate $k_{\text{cat}}=0.4$ demonstrate a delayed arrangement around four non-overlapping asters (Fig.~\ref{fig:7}D) resembling the fixed-length-microtubule network (Fig.~\ref{fig:6}A and Fig.~\ref{fig:7}B)). Longer fibers at lower catastrophe rates establish links between distant clusters of the networks, thus supporting the more effective accumulation of the fibers around a single giant cluster.

\nik{Considering a more realistic value of the unloaded motor speed $v_0=-0.1$ $\mu$m/s at a 4-fold increased simulation time, we show that the microtubule-motor mixture immediately self-organizes into a single aster (Fig.~\ref{fig:B3}A). It contrasts with the previous multi-aster patterns at higher motor speed $v_0=-0.8$ $\mu$m/s. Apart from that, the hysteresis of microtubule self-organization was preserved at a lower motor speed -- the aster formed during the forward transition remained ``frozen'' during the backward transition. The presence of hysteresis was also verified by evaluating the macroscopic quantifiers of the microtubule steady-state (Fig.~\ref{fig:B3}B-D). Based on this observation, we conclude that (i) a higher motor speed promotes local spatial patterning, resulting in the formation of multiple asters, while a slower motor speed promotes global patterning; and (ii) hysteresis is a generic property of the dynein-driven microtubule self-organization and is not affected by the motor speed.}

Taken together, our data suggest that the hysteresis of self-organization in microtubule-dynein mixture is robust against the variation of considered parameters of motors and microtubules, namely dynein unbinding rate $p_0$ and microtubule catastrophe rate $k_{\text{cat}}$. Higher unbinding and lower catastrophe rates ensured the most efficient percolation transition.

\section{Conclusion}
\label{sec:Concl}

In this computational study, we addressed the properties of radial arrangement of the microtubules -- asters -- driven by molecular motors. We started our analysis by considering a well-studied minimal model of fixed-length microtubules and dynein proteins capable of aster formation. We proposed an entropy-based measure to evaluate the system's steady-state's spatial (dis)order and tested it on a minimal model. A microtubule-motor mixture at low concentrations of molecular motors represented an almost non-contractile network -- a pattern of low spatial order according to entropy estimates. Such a pattern consisted of multiple poorly concentrated, sparsely connected, and irregularly positioned asters. With increasing concentration of motor proteins, the system switched to more ordered states, e.g., rectangular or triangular compositions of stronger nucleated asters, exhibiting lower entropy. Finally, at a high concentration of dynein, the system reached a state of minimal complexity -- a single giant aster.

Further analysis revealed that the dynein-driven microtubule motility changed smoothly under the variation of the dynein concentration, potentially explaining a similar smooth variation of the steady-state entropy but the number of asters undergoing a cascade of discontinuous transitions. Representing a dynein-fiber network as a graph by considering fibers as edges and their respective mechanical connections through dyneins as edges, we found that the median fibers' degree at the beginning of self-organization undergoes similar discontinuous transitions as the number of asters. Moreover, we analyzed the number of clusters and the size of the largest cluster in the microtubule-motor network. Variations of these quantifiers under a growing concentration of dyneins resembled the percolation transition to a giant cluster, and the onset of the percolation transition matched the increase in the median microtubule network degree. A network-based analysis established an association between the steady-state microtubule pattern's spatial entropy and the dynein-fiber network's median degree. We found a reasonable clustering of the steady states on the entropy-degree plane such that one can distinguish between (i) non-contractile networks dominated by the isolated pairs of sparsely connected fibers, (ii) states with multiple spatial scales dominated by the locally connected microtubule communities, and (iii) states with single spatial scale dominated by the microtubules establishing coupling of a higher order.

Finally, we established that the dynein-driven arrangement of the microtubules exhibits hysteresis under the variation of dynein concentration. During the forward continuation, i.e., a progressive increase in dyneins, the microtubule-motor mixture undergoes a familiar percolation transition from a disordered non-contractile network to several well-organized and highly concentrated asters. Backward continuation, i.e., a progressive decrease in dynein, did not essentially disrupt the pre-formed asters. This result suggests that dyneins robustly maintain the radial arrangement of the microtubules once concentrated around the aster center. The revealed property of aster formation is pivotal for the stability and reproducible organization of meiotic and mitotic spindles, of which the asters are critical structural elements.

To conclude, our study uses a complexity science approach to describe and explore the self-organization of microtubules driven by interaction with molecular motors. It complements existing works and offers new insights into phase transitions in microtubule-motor mixtures. Prospectively, it unlocks a range of intriguing open questions, e.g., how the transition to steady states and their (dis)order would change under (i) the interaction with other types of motor proteins (a mixture of motor proteins of different types); (ii) mechanical interaction with other filament networks (F-actins)~\cite{pelletier2020co}; (iii) positioning of the microtubule asters and nuclei~\cite{sami2022dynein}.

\section*{Declaration of Competing Interest}
The authors declare that they have no known competing financial interests or personal relationships that could have appeared to influence the work reported in this paper.

\section*{Acknowledgments}
D.R-R. acknowledges support by the internal funds KU Leuven (grant no. PDM/20/153), the Ministry of Universities through the ``Pla de Recuperació, Transformació i Resilència'', and by the EU (NextGenerationEU) together with the Universitat de les Illes Balears. LG acknowledges financial support by the Research-Foundation Flanders (FWO-Vlaanderen) (grant no. G074321N). We also thank Felix E. Nolet for the valuable discussions on related topics leading up to this study. 


\bibliographystyle{elsarticle-num}
%\bibliographystyle{cas-model2-names}

% \bibliography{references}

\newpage
\appendix
\setcounter{figure}{0}
\section{Setting up simulations in the Cytosim}
\label{app:A}
\renewcommand\thetable{A\arabic{table}}

To reproduce the results presented in the paper, one should install the Cytosim software and set up the configuration file following the instructions provided on the web page of this project~\cite{gitlabCytosim}. The configuration file contains the list of parameters of the simulation engine and the system to be modeled. The tables below outline the model parameters used to simulate the microtubule-motor interaction in the case of fixed-length microtubules (Table~\ref{tab: parameters_static}), large chamber (Table~\ref{tab: parameters_large}), and dynamic instability of the microtubules (Table~\ref{tab: parameters_dynamic}).

\begin{table}[H]
\caption{\label{tab: parameters_static} Parameters used to configure a numerical model of the fixed-length-microtubule-motor interaction in the Cytosim}
    \begin{tabular}{ll}
        \textbf{Model parameter} & \textbf{Value}\\
        \hline
        \hline
        \\
        \textbf{Cytoplasm} &  \\
        Viscosity & 0.05 pNs/$\mu$m$^2$ \\
        Chamber size, $l_{\text{ch}}$ &  50 $\mu$m \\
        \hline
         \\
        \textbf{Microtubules} & \\
        Number of fibers, $N_{\text{MT}}$ & 2000\\
        Fiber length, $l$ & 10 $\mu$m\\
        Rod segment length, $l_{\text{seg}}$ & 0.5 $\mu$m\\
        \hline
        \\
        \textbf{Motor protein complexes (dyneins)} & \\
        Binding rate, $p_{\text{bind}}$ & 10 s$^{-1}$\\
        Binding range, $\varepsilon$ & 0.01 $\mu$m \\
        Unloaded unbinding rate, $p_{0}$ & [0.05,0.1,0.2] s$^{-1}$\\
        Unbinding force, $f_{\text{unbind}}$ & 3 pN\\
        Unloaded dynein velocity, $v_{0}$ & $-$0.8 $\mu$m/s\\
        Stall force, $f_{\text{stall}}$ & 5 pN\\
        Number of dynein complexes, $N_{\text{mot}}$ & [1000,5000]
    \end{tabular}
\end{table}

\begin{table}[H]
\caption{\label{tab: parameters_large} Parameters used to configure a numerical model of the fixed-length-microtubule-motor interaction in a larger chamber in the Cytosim}
    \begin{tabular}{ll}
        \textbf{Model parameter} & \textbf{Value}\\
        \hline
        \hline
        \\
        \textbf{Cytoplasm} &  \\
        Viscosity & 0.05 pNs/$\mu$m$^2$ \\
        Chamber size, $l_{\text{ch}}$ &  75 $\mu$m \\
        \hline
         \\
        \textbf{Microtubules} & \\
        Number of fibers, $N_{\text{MT}}$ & 4500\\
        Fiber length, $l$ & 10 $\mu$m\\
        Rod segment length, $l_{\text{seg}}$ & 0.5 $\mu$m\\
        \hline
        \\
        \textbf{Motor protein complexes (dyneins)} & \\
        Binding rate, $p_{\text{bind}}$ & 10 s$^{-1}$\\
        Binding range, $\varepsilon$ & 0.01 $\mu$m \\
        Unloaded unbinding rate, $p_{0}$ & 0.1 s$^{-1}$\\
        Unbinding force, $f_{\text{unbind}}$ & 3 pN\\
        Unloaded dynein velocity, $v_{0}$ & $-$0.8 $\mu$m/s\\
        Stall force, $f_{\text{stall}}$ & 5 pN\\
        Number of dynein complexes, $N_{\text{mot}}$ & [2250,11250]
    \end{tabular}
\end{table}

\begin{table}[H]
\caption{\label{tab: parameters_dynamic} Parameters used to configure a numerical model of the microtubule-motor interaction in the Cytosim taking into account dynamic instability}
    \begin{tabular}{ll}
        \textbf{Model parameter} & \textbf{Value}\\
        \hline
        \hline
        \\
        \textbf{Cytoplasm} &  \\
        Viscosity & 0.05 pNs/$\mu$m$^2$ \\
        Chamber size, $l_{\text{ch}}$ &  50 $\mu$m \\
        \hline
         \\
        \textbf{Microtubules} & \\
        Number of fibers, $N_{\text{MT}}$ & 2000\\
        Growth speed, $v_g$ & 1 $\mu$m/s\\
        Shrinkage speed, $v_s$ & -10 $\mu$m/s\\
        Catastrophe rate, $k_{\text{cat}}$ & [0.2,0.4] s$^{-1}$\\
        Rescue rate, $k_{\text{res}}$ & 0.0 s$^{-1}$\\
        Minimal length, $l_{\text{min}}$ & 0.02 $\mu$m \\
        Rod segment length, $l_{\text{seg}}$ & 0.5 $\mu$m\\
        \hline
        \\
        \textbf{Motor protein complexes (dyneins)} & \\
        Binding rate, $p_{\text{bind}}$ & 10 s$^{-1}$\\
        Binding range, $\varepsilon$ & 0.01 $\mu$m \\
        Unloaded unbinding rate, $p_{0}$ & 0.1 s$^{-1}$\\
        Unbinding force, $f_{\text{unbind}}$ & 3 pN\\
        Unloaded dynein velocity, $v_{0}$ & $-$0.8 $\mu$m/s\\
        Stall force, $f_{\text{stall}}$ & 5 pN\\
        Number of dynein complexes, $N_{\text{mot}}$ & [1000,5000]
    \end{tabular}
\end{table}

\newpage
\section{MT self-organization in a larger cell}
\renewcommand\thefigure{B\arabic{figure}}
\label{app:B}

\begin{figure*}[!t]
	\centerline{\includegraphics[scale=0.20]{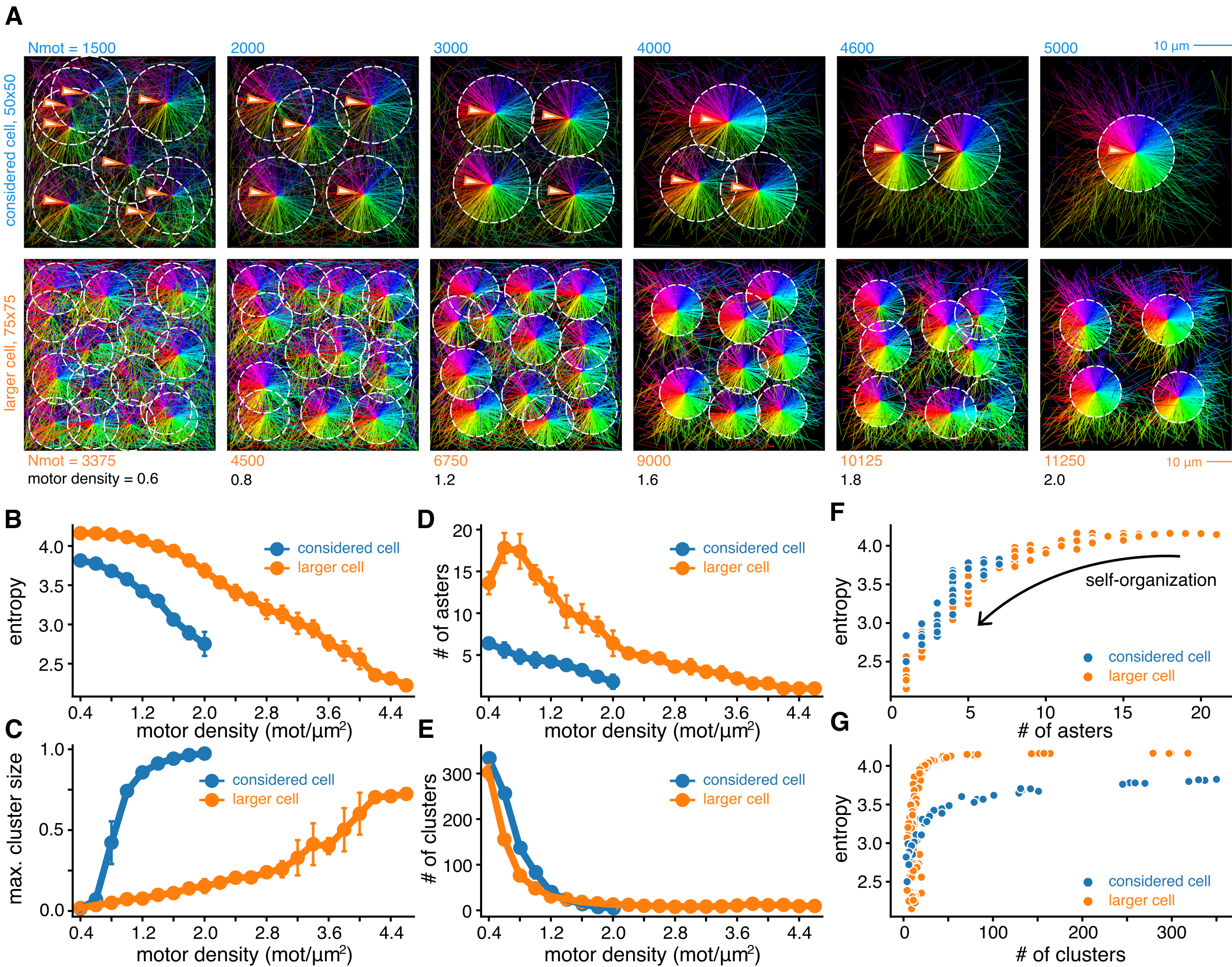}}
	\caption{\textbf{The effect of the chamber size on the microtubule-motor steady-state. A.} Snapshots of the microtubule-motor steady-state in a considered chamber (top row) and a chamber of a larger size (bottom row) at the same motor concentration. \textbf{B.} Entropy of the steady-state under the variation of the motor concentration (mean$\pm$SD across $n=5$ realizations). \textbf{C.} Maximal cluster size in the steady-state under the variation of the motor concentration (mean$\pm$SD across $n=5$ realizations). \textbf{D.} Number of asters in the steady-state under the variation of the motor concentration (median and IQR across $n=5$ realizations). \textbf{E.} Number of steady-state clusters under the motor concentration variation (mean$\pm$SD across $n=5$ realizations). \textbf{F.} Entropy versus the number of asters. \textbf{G.} Entropy versus the number of clusters.}\label{fig:B1}
\end{figure*}

\begin{figure*}[!t]
	\centerline{\includegraphics[scale=0.20]{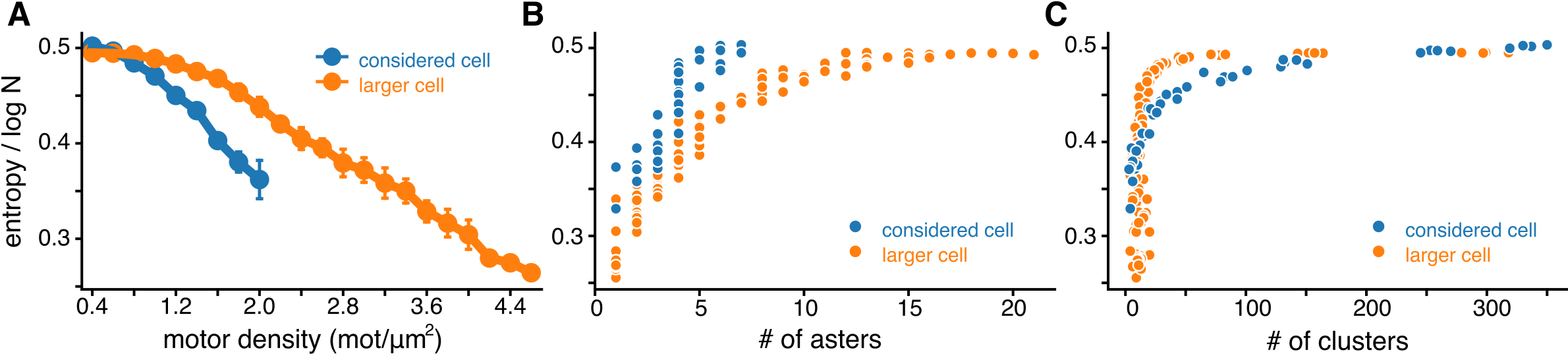}}
	\caption{\textbf{The effect of the chamber size assessed via normalized entropy. A.} Normalized steady-state entropy under the motor concentration variation (mean$\pm$SD across $n=5$ realizations). \textbf{B.} Normalized entropy versus the number of asters. \textbf{G.} Normalized entropy versus the number of clusters.}\label{fig:B2}
\end{figure*}

\begin{figure*}[!t]
	\centerline{\includegraphics[scale=0.20]{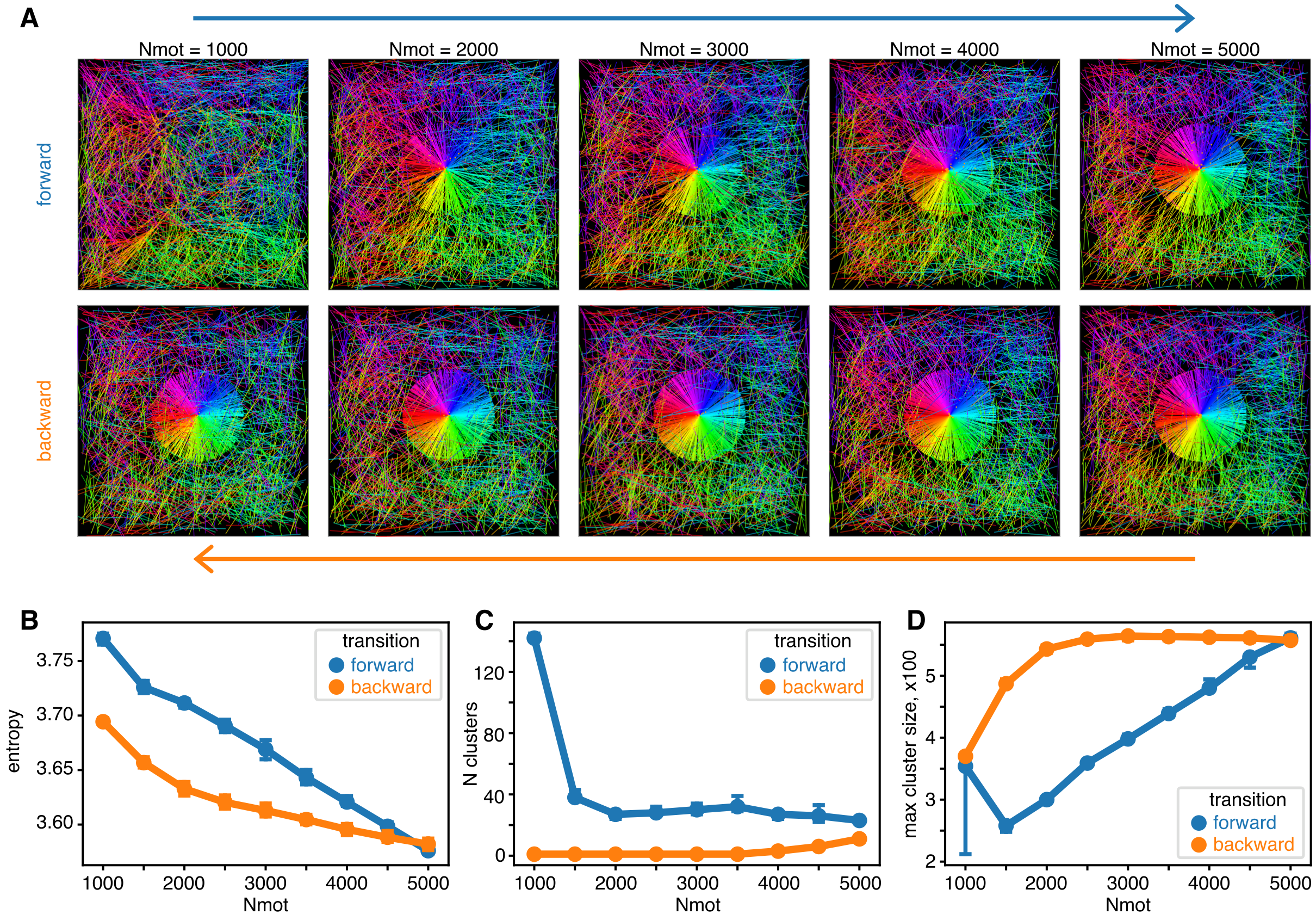}}
	\caption{\textbf{Hysteresis in microtubule self-organization is still present at a realistic dynein velocity (-0.1 $\mu$m/s). A.} Evolution of a steady state in a microtubule-motor mixture. Top row: forward transition, i.e., continuously adding new dynein complexes into the chamber; bottom row: backward transition, i.e., continuously removing existing dynein complexes from the chamber. \textbf{B-D.} Dependence of the steady state quantifiers on the number of dynein complexes during forward (blue) and backward (orange) transitions: (\textbf{B}) entropy (mean$\pm$SD across $n=5$ realizations); (\textbf{C}) number of clusters (median and IQR across $n=5$ realizations ); (\textbf{D}) maximal cluster size (median and IQR across $n=5$ realizations).}\label{fig:B3}
\end{figure*}


\end{document}